\documentstyle[aps,prc,twocolumn,floats,epsfig]{revtex}

\draft

\begin{document}

\twocolumn[\columnwidth\textwidth\csname@twocolumnfalse\endcsname

\author{A.T. Kruppa,$^{1,2}$ 
        M. Bender,$^{3-5}$
        W. Nazarewicz,$^{3,5,6}$  
        P.-G. Reinhard,$^{1,7}$
        T. Vertse,$^{1,2}$
        S. {\'C}wiok$^{8}$
}

\address{$^1$Joint Institute for Heavy Ion Research,
             Oak Ridge National Laboratory,
             P.O. Box 2008, Oak Ridge,   Tennessee 37831}
\address{$^2$Institute of Nuclear Research 
             of the Hungarian Academy of Sciences,
             H-4001 Debrecen, Pf. 51, Hungary}
\address{$^3$Department of Physics and Astronomy, University
             of Tennessee, Knoxville,  Tennessee 37996}
\address{$^4$Department of Physics and Astronomy, University
             of North Carolina, Chapel Hill, North Carolina  27599}
\address{$^5$Physics Division, Oak Ridge National Laboratory,
             P.O. Box 2008, Oak Ridge,   Tennessee 37831}
\address{$^6$Institute of Theoretical Physics, Warsaw University,
             ul. Ho\.za 69, PL-00681, Warsaw, Poland}
\address{$^7$Institut f\"ur Theoretische Physik II, Universit\"at Erlangen
             Staudtstr.\ 7, D-91058 Erlangen, Germany}       
\address{$^8$Faculty  of Physics, Warsaw University of Technology,
             ul. Koszykowa 75, PL-00662, Warsaw, Poland}

\title{Shell Corrections of Superheavy Nuclei 
       in Self-Consistent Calculations}

\date{October 15, 1999}

\maketitle

\begin{abstract}
Shell corrections to the nuclear binding energy
as a measure of shell effects in superheavy nuclei 
are studied within the self-consistent 
Skyrme-Hartree-Fock and Relativistic Mean-Field theories. 
Due to the presence of low-lying proton
continuum resulting in a free particle gas, 
special attention is paid to the treatment of
single-particle level density. To cure the pathological 
behavior of shell correction
around the particle threshold, the method based on the 
Green's function approach has been
adopted. It is demonstrated that for the vast majority 
of Skyrme interactions commonly
employed in nuclear structure calculations, the
 strongest shell stabilization appears for
$Z$=124, and 126, and for $N$=184. On the other 
hand, in the relativistic approaches
the strongest spherical shell effect appears systematically
for $Z$=120 and $N$=172. This difference has probably
its roots in  the spin-orbit potential.
We have also shown that, in contrast to shell
corrections  which are fairly independent on the force,
macroscopic energies extracted from self-consistent calculations 
strongly depend on the actual force parametrisation used. 
That is, the $A$ and $Z$ dependence of mass
surface when extrapolating to unknown superheavy nuclei is prone 
to  significant theoretical uncertainties.
\end{abstract}

\pacs{PACS number(s): 21.10.Dr, 21.10.Pc, 21.60.Jz, 27.90.+b}

\addvspace{5mm}]

\narrowtext

\section{Introduction}

\noindent
The stability of the heaviest and superheavy elements has 
been a long-standing fundamental question in nuclear science.
Theoretically, the mere existence of the heaviest elements 
with $Z$$>$104 is entirely due to quantal shell effects. Indeed,
for these nuclei
the shape of the classical nuclear  droplet, governed by surface tension and 
Coulomb repulsion, is unstable to surface distortions
driving these nuclei to spontaneous fission. That is, if the heaviest
nuclei  were governed by the classical liquid drop model, they would
fission immediately  from their ground states
due to the large electric charge. However, in the mid-sixties,
with the invention of the shell-correction method,
it was realized that  long-lived superheavy elements (SHE) 
with very large  atomic numbers 
could exist due to the strong shell stabilization
\cite{[Mye66],[Sob66],[Mel67],[Mos69]}. 
%
%
\begin{figure*}[t!]
\begin{center}
\leavevmode
\epsfxsize=16cm
\epsfbox{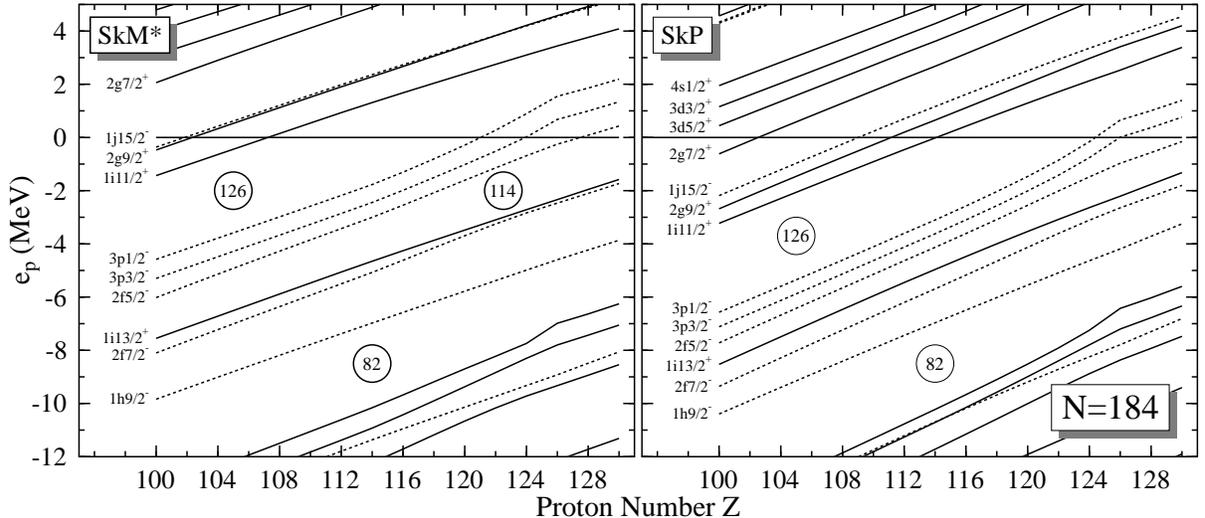}
\end{center}
\caption{Single-proton levels for $N$=184 isotones with
110$\le$$Z$$\le$130 calculated in the Skyrme-Hartree-Fock
model with SkM$^*$ (left) and SkP effective interactions.
Positive (negative) parity levels are indicated by solid (dashed)
lines and by their spherical labels ($nlj$).
Note that in both cases the nucleus $Z$=126 is proton-unbound,
i.e., the $p_{1/2}$ shell has positive energy. 
}
\label{spe1}
\end{figure*}
%
%

In spite of tremendous experimental
effort, after about thirty years of the quest for  superheavy
elements,  the borders of the upper-right end of
the nuclear chart are still unknown \cite{[Hof99]}.
However, it has to be emphasized that the recent
years  also brought significant  progress  in the production
of the heaviest nuclei \cite{[Hof99],[Hof97a]}.
During 1995-96,  three new elements, $Z$=110, 111, and 112,
were synthesized by means of both cold and hot fusion reactions
\cite{[Hof95],[Ghi95],[Laz96],[Hof96]}.
These heaviest  isotopes  decay predominantly by groups of
$\alpha$ particles ($\alpha$ chains) as expected theoretically 
\cite{[Cwi87a],[Smo97],[Smo99]}.  
Recently, two stunning discoveries have been made.
Firstly,  hot fusion experiments performed in Dubna employing
$^{48}$Ca+$^{244}$Pu and $^{48}$Ca+$^{242}$Pu ``hot fusion"
 reactions \cite{[Oga99]} gave evidence for 
the synthesis of two isotopes ($A$=287 and 289) of the  element $Z$=114.
Secondly, the Berkeley-Oregon  team, utilizing the ``cold fusion"
reaction $^{86}$Kr+$^{208}$Pb \cite{[Nin99]}, observed three 
$\alpha$-decay chains attributed to the decay of the new element 
$Z$=118, $A$=293. The measured $\alpha$-decay chains
$^{289}$114 and $^{293}$118 turned out to be consistent
with predictions of the Skyrme-Hartree-Fock (SHF)  theory \cite{[Cwi99]}
and the Relativistic Mean-Field (RMF) theory  \cite{[Ben99]}.

The goal of the present work is to study shell closures in SHE. 
To that end we use as a tool microscopic shell 
corrections extracted from self-consistent calculations.
For medium-mass and heavy nuclei, self-consistent mean-field theory is a very
useful starting point \cite{[Abe90a]}. Nowadays, SHF and RMF calculations
with realistic effective forces are able to describe
 global nuclear 
properties  with an accuracy which is 
comparable to that obtained in more phenomenological 
macroscopic-microscopic models based on the shell-correction method. 

In previous work \cite{[Cwi96]}, shell energies for SHE elements 
were extracted by subtracting from calculated HF 
binding energies the macroscopic Yukawa-plus-exponential 
mass formula \cite{[Mol88]} with parameters of 
Ref.~\cite{[Cwi94a]}. In another work, based on the 
RMF theory \cite{[Lal96]}, shell corrections were extracted for
the heaviest deformed nuclei using the standard Strutinsky method 
in which the positive-energy spectrum was  approximated by 
quasi-bound states. Neither procedure can be considered as 
satisfactory. A proper treatment of continuum states is achieved
with a Green's function method \cite{[Kru98]}. We employ this 
method for the present study of shell corrections of SHE.

The material contained in this study is organized as follows.
The motivation of this work is outlined in Sect.~\ref{motiv},
Section~\ref{SET} contains a brief discussion of the Strutinsky energy theorem
on which the concept of shell correction is based. The 
Green's Function HF  method used to extract the single-particle level density
is presented in Sect.~\ref{GFHF}. Section~\ref{models} discusses the 
details of our HF and RMF models and describes the Strutinsky procedure
employed. The results of calculations
for shell corrections in spherical SHE and for macroscopic
energies extracted from self-consistent binding energies
are discussed  in Sec.~\ref{results}. Finally, Sec.~\ref{conclusions} contains
the main conclusions of this work.

\section{Motivation}\label{motiv}

\noindent
All the heaviest elements found recently
are believed to be well deformed.
Indeed, the measured $\alpha$-decay energies,  along with
complementary syntheses of new neutron-rich isotopes
of elements $Z$=106 and $Z$=108, have furnished confirmation of
the special stability of the deformed
shell at $N$=162 predicted by theory \cite{[Cwi83],[Mol94a]}.
Beautiful  experimental confirmation  of large quadrupole
deformations in this mass region comes from gamma-ray
spectroscopy. Recent experimental works
\cite{[Rei99],[Lei99]} succeeded in
identifying the ground-state band of $^{254}$No (the
heaviest nucleus studied in gamma-ray spectroscopy so far).
The quadrupole deformation of $^{254}$No,
inferred from the energy of the deduced $2^+$ state,  is
in  nice agreement with theoretical predictions 
\cite{[Cwi96],[Cwi94a],[Pat91],[Bur98]}. Still heavier and 
more neutron-rich elements are expected to be spherical  
due to the proximity of the neutron shell at $N$=184.
This is the region of SHE which we will investigate here.
%
%
\begin{figure*}[t!]
\begin{center}
\leavevmode
\epsfxsize=16cm
\epsfbox{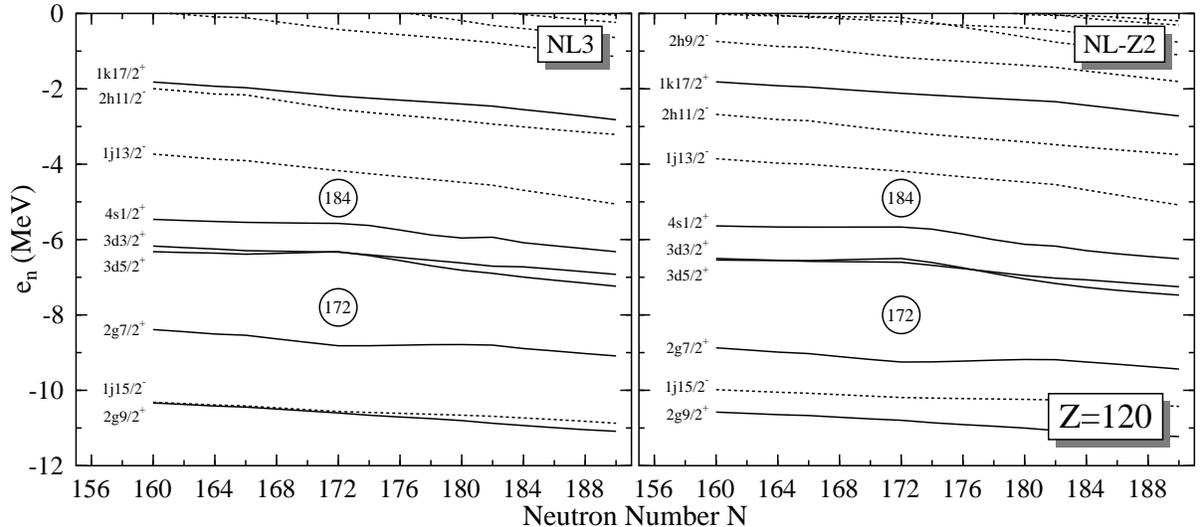}
\end{center}
\caption{Single-neutron levels for $Z$=120 isotopes with
160$\le$$N$$\le$190 calculated in the RMF approach
with NL3 (left) and NL-Z2 non-linear parametrizations.
The line convention is the same as in Fig.~\protect\ref{spe1}.
Note large neutron gaps at $N$=172 and 184. 
}
\label{spe2}
\end{figure*}
%
%

In spite of an impressive agreement with available experimental data for
the heaviest elements, theoretical uncertainties 
are large when extrapolating to unknown nuclei with greater atomic numbers.
As discussed in Refs.\  \cite{[Cwi96],[Ben99a]}, 
the main factors that influence the single-proton
shell structure of SHE are (i) the Coulomb potential and (ii) the spin-orbit
splitting. As far as the protons are concerned, the important spherical shells
are the closely spaced  $1i_{13/2}$ and  $2f_{7/2}$ levels which
appear just below the $Z$=114 gap, the   
$2f_{5/2}$ shell which becomes occupied at $Z$=120, the
$3p_{3/2}$ shell which becomes occupied at $Z$=124, and the 
$3p_{1/2}$ and $1i_{11/2}$ orbitals whose splitting determines
the size of the $Z$=126 magic gap. Interestingly, while the {\em ordering}
of single-proton states is practically the same for all the self-consistent
approaches with realistic effective interactions
(see Fig.~\ref{spe1} and single-particle
diagrams in Refs.~\cite{[Cwi96],[Ben99a]}), their relative positions
vary depending on the choice of
force parameters. Since in the region of SHE the single-particle
level density is relatively large, small shifts in positions of
single-particle levels can influence the strength of  single-particle
gaps and be crucial for determining the shell stability of a nucleus.
As a result,
there is no consensus between theorists concerning the next proton magic
gap beyond $Z$=82. While most 
macroscopic-microscopic  (non-self-consistent) approaches predict 
$Z$=114 to be magic, self-consistent calculations suggest that 
the center of the proton shell stability should be moved up to higher proton numbers,
$Z$=120, 124 or 126 \cite{[Cwi96],[Bur98],[Ben99a],[Rut97]}. It is to be noted
that the Coulomb potential mainly influences the magnitude of the $Z$=114
gap. (Here, the self-consistent treatment of Coulomb energy is a key factor.)
On the other hand, the
spin-orbit interaction determines the  position of the
$2f$ and $3p$ shells which define the proton shell structure above $Z$$>$114.

The spherical neutron shell structure is governed by the following
orbitals: $1j_{15/2}$ (below the $N$=164 gap), $2g_{7/2}$, $3d_{5/2}$,
$3d_{3/2}$, and $4s_{1/2}$ and $1j_{13/2}$ whose splitting
determines the size of the $N$=184 spherical gap (see Fig.~\ref{spe2}
and Refs.~\cite{[Cwi96],[Ben99a]}). Again, similar to
the proton case, the order of the single-neutron orbitals between
$N$=164 and 184 is rather
robust, while sizes of single-particle gaps vary. For instance,
the $N$=172 gap, predicted by the RMF calculations shown in Fig.~\ref{spe2},
results from the large energy splitting between the $2g_{7/2}$
and  $3d_{5/2}$ shells. In non-relativistic models, these two orbitals
are very close in energy, and this degeneracy is related to the
pseudo-spin symmetry \cite{[Ari69],[Hec69]}. Interestingly,
in the SHF calculations, the pseudo-spin degeneracy holds in most cases. Namely,
certain neutron orbitals group in pairs (pseudo-spin doublets):
($2g_{7/2}$, $3d_{5/2}$), ($3d_{3/2}$, $4s_{1/2}$), and the same holds for 
proton orbitals, e.g., ($2f_{5/2}$, $3p_{3/2}$). 
Considering the fact that the idea of pseudo-spin 
has relativistic roots \cite{[Gin97],[Gin98]}, it is surprising 
to see that this symmetry is so dramatically 
violated  in the RMF theory. As a matter of fact,
the presence of pronounced magic gaps at $Z$=120 and 
$N$=172 in RMF models (see below) is a direct manifestation of 
the pseudo-spin symmetry breaking.

As discussed in Ref.~\cite{[Cwi96]}, neutron-deficient
superheavy nuclei are expected to be unstable to proton emission. 
Indeed, as seen in Fig.~\ref{spe1}, the proton  $3p_{1/2}$ shell 
has positive energy for $Z$$\ge$126, i.e., 
in these nuclei the $3p_{1/2}$ level is  a narrow resonance.
Due to huge Coulomb barriers, superheavy nuclei with $Q_p$$<$1.5\,MeV
are practically proton-stable \cite{[Cwi96]}. However,
the higher-lying single-proton orbitals are expected to have sizable 
proton widths.

In order to assess the magnitude of shell effects determined by the bunchiness
of single-particle levels, it is useful to apply the Strutinsky renormalization
procedure \cite{[Str67],[Str68],[Bra72]} which 
makes it possible  to calculate the shell correction energy.
Unfortunately, the standard way of extracting shell correction breaks down
for weakly bound nuclei where the contribution from the particle continuum
becomes important \cite{[Naz94]}. Recently, a new method of calculating
shell correction, based on the correct treatment of resonances,
has been developed \cite{[San97a],[Ver98]}. The improved method is based on
the theory of Gamow states (eigenstates of one-body Hamiltonian
with purely outgoing boundary conditions) which can be calculated numerically
for commonly used optical-model potentials \cite{[Ver82]}. 
While this ``exact" procedure cannot be easily adopted to the
case of microscopic  self-consistent  potentials, its simplified version
applying the Green's-function method can \cite{[Kru98]}.

\section{Shell Correction and the Energy Theorem}\label{SET}

\noindent
The main assumption of the shell-correction (macro\-sco\-pic-microscopic) 
method \cite{[Str67],[Str68],[Bra72],[Bun72]} is that the total energy
of a nucleus can be decomposed  into two parts:
\begin{equation}
\label{Eshell}
E 
= \tilde{E}+E_{\rm shell},
\end{equation}
where $\tilde{E}$ is the macroscopic energy (smoothly depending on the
number of
nucleons and thus associated with the ``uniform" distribution of
single-particle orbitals) and $E_{\rm shell}$ is the shell-correction term
that fluctuates with particle number reflecting the non-uniformities 
(bunchiness) of the single-particle level distribution.
In order to make a separation (\ref{Eshell}), one starts from
the one-body HF  density matrix $\rho$
\begin{equation}\label{rhosp}
{\rho}(\bbox{r}',\bbox{r}) = \sum_{i}
n_i \phi_i(\bbox{r}')\phi^*_i(\bbox{r}),
\end{equation}
which can be decomposed into a ``smoothed" density $\tilde \rho$
and a correction $\delta\rho$, which fluctuates with the shell filling
\begin{equation}\label{rhot}
\rho =\tilde\rho + \delta\rho.
\end{equation}
In Eq.~(\ref{rhosp}), $n_i$ is the single-particle occupation
coefficient which is equal to 1(0) if the level $e_i$ is occupied
(empty).
The smoothed single-particle density $\tilde\rho$ can be 
expressed by means of the smoothed distribution numbers 
$\tilde{n}_i$ \cite{[Bra74a]}:
\begin{equation}\label{rhos}
{\tilde\rho}(\bbox{r}',\bbox{r}) = \sum_{i}\tilde{n}_i\phi_i(\bbox{r}')\phi^*_i(\bbox{r}).
\end{equation}
When considered as a function of the single-particle energies $e_i$,
the numbers $\tilde{n}_i$ vary smoothly in an
energy interval of the order of the energy difference between major shells.
The averaged HF Hamiltonian $\tilde h_{\rm HF}$ can be directly obtained 
from $\tilde{\rho}$. The expectation value of a HF Hamiltonian
(containing the kinetic energy, $t$ and the two-body
interaction, $\bar v$) can then be written in terms of
$\tilde\rho$ and $\delta\rho$ \cite{[Bun72],[RS80]}:
\begin{equation}
\label{O}
E_{\rm HF} 
= {\rm Tr}(t\rho) 
  + {1\over 2}{\rm Tr\,Tr}(\rho\bar v\rho) 
= \tilde E + E_{\rm osc} + O(\delta\rho^2),
\end{equation}
where
\begin{equation}
\tilde E 
= {\rm Tr} (t\tilde\rho) 
  + {1\over 2}{\rm Tr\,Tr}(\tilde\rho\bar v\tilde\rho)
\end{equation}
is the average  part of $E_{\rm HF}$ and
\begin{equation}\label{Eosc}
E_{\rm osc} = {\rm Tr}(\tilde h_{\rm HF}\delta\rho)
\quad \mbox{with} \quad
\tilde{h}_{\rm HF} = t+{\rm Tr}(\bar v\tilde\rho)
\end{equation}
is the first-order term in $\delta\rho$ representing the
shell-correction contribution to $E_{\rm HF}$.
If a  deformed phenomenological  potential
gives a similar spectrum to the averaged HF potential
$\tilde h_{\rm HF}$, then the oscillatory part of $E_{\rm HF}$, given by
Eq.~(\ref{Eosc}), is very close to that of the deformed shell model,
$E_{\rm shell}$=$E_{\rm osc}+O(\delta \rho^2)$. The second-order term
in Eq.~(\ref{O}) is usually very small and can be neglected \cite{[Bra81]}.
The above relation, known as the {\em Strutinsky Energy Theorem},
makes it possible to calculate the total energy using the
non-self-consistent, deformed independent-particle model;
the average part $\tilde E$ is usually replaced by the corresponding
phenomenological liquid-drop
(or droplet) model value, $E_{\rm macro}$.
It is important that $E_{\rm shell}$ must not contain any regular
(smooth) terms analogous
to those already included in the phenomenological macroscopic part.
The numerical proof of the Energy Theorem was carried out
 by Brack and Quentin
\cite{[Bra75]} who demonstrated that Eq.~(\ref{Eshell}) holds
for  $E_{\rm shell}$  defined by means of  the smoothed single-particle
energies (eigenvalues of  $\tilde h_{\rm HF}$).

In this work, we use a simpler expression to extract the
shell correction from the HF binding energy, which should 
also be accurate up to $O(\delta\rho^2)$. Namely, as an 
input to the Strutinsky procedure we take the self-consistent
single-particle HF energies, $e^{\rm HF}_i$. In this case,
the shell correction is given by
\begin{equation}\label{esp}
{E}_{\rm shell}(\rho) 
= \sum_i ( n_i - \tilde{n}_i) e_i + O (\delta \rho^2).
\end{equation}
The equivalent macroscopic energy can easily be computed by
taking the difference
\begin{equation}\label{Emacro}
E_{\rm macro} \approx \tilde{E}^{\rm HF} 
= E(\rho) - {E}_{\rm shell}(\rho).
\end{equation}

\section{Green's Function Hartree-Fock Approach to Shell Correction}\label{GFHF}

\noindent
The HF equation is generally solved using a  harmonic oscillator 
expansion method or by means of a discretization in a three-dimensional
box. In both cases, a great number of 
unphysical states with positive energy appear. The effect of 
these quasi-bound states is disastrous for the Strutinsky 
renormalization procedure 
\cite{[Kru98],[Naz94],[San97a],[Ver98],[Bol72]}.
Indeed, if one smoothes out the single-particle energy density,
\begin{equation}\label{gsp}
g_{\rm sp}(e)=\sum_i \delta\left(e-e_i^{\rm HF}\right),
\end{equation}
it would diverge at zero energy 
because the presence of the unphysical positive 
energy states. Consequently,  the resulting
shell correction  becomes unreliable.

In order to avoid the divergence of $g(e)$ around the threshold,
we apply the 
Green's-function method \cite{[Kru98],[Bal70],[Bal71],[Shl92],[Shl96]} 
for the calculation of the single-particle level density.
In this method, the level density is  given by the expression 
\begin{equation}\label{GFg}
g(e)
=-{1\over \pi}\Im
  \left\{ {\rm Tr} \left[ \hat{G}^+(e)
                         -\hat{G}^+_{\rm free}(e)
                   \right]
  \right\},
\end{equation}
where $\hat{G}^+(e)=(e-\hat{h}+i0)^{-1}$ is the 
outgoing Green's operator of the single-particle
Hamiltonian $\hat{h}(\rho)$, and $\hat{G}^+_{\rm free}$ is the 
free outgoing Green's operator that belongs to
the ``free'' single-particle Hamiltonian. This latter 
is derived from the full HF Hamiltonian in such a way that those terms are
kept which are related to the kinetic energy density and to  the direct 
Coulomb term. The interpretation of Eq.~(\ref{GFg}) is straightforward:
the second term in Eq.~(\ref{GFg}) contains the contribution to the
single-particle level density originating from the gas of free particles.
 
The single-particle  level density  defined by the 
Green's-function expression (\ref{GFg})
behaves smoothly around the zero-energy threshold;
for finite-depth Hamiltonians  this
definition is the only meaningful way of introducing $g(e)$.
The level density (\ref{GFg}) automatically takes into 
account the effect of the particle continuum  which may influence 
the results of shell-correction calculations 
\cite{[San97a],[Ver98]}, especially pronounced
 for systems where the Fermi level 
is close to zero, i.e., drip-line nuclei.

Because it is difficult to calculate the Green's-function,
in this work  we applied 
the approximation introduced in Ref.~\cite{[Kru98]}. In this approach, 
the single-particle level density 
is expressed as
\begin{equation}\label{gapp}
g(e)\approx\sum_i \delta\left(e-e^{\rm HF}_i\right)
-\sum_i \delta\left(e-e^{\rm HF,free}_i\right),
\end{equation} 
where $e^{\rm HF,free}_i$ are the eigenvalues of the free  
one-body HF Hamiltonian. As usual in the Strutinsky procedure,
the smooth level density can be
obtained by folding $g(e)$ with a smoothing function
$f(x)$:
\begin{eqnarray}\label{ggsmooth}
\tilde{g}(e) 
& = & {1\over\gamma} \int_{-\infty}^{+\infty} de'~ g(e') \; 
      f\left(\frac{e'-e}{\gamma}\right)
      \nonumber \\
& = & \tilde{g}_0(e) - \tilde{g}_{\rm free}(e),
\end{eqnarray}
where $\gamma$ is the smoothing width, 
$\tilde{g}_0(e)$ is the  smooth level density obtained from the
HF spectrum (including the quasi-bound states), and
$\tilde{g}_{\rm free}(e)$ is the contribution to the
smooth level density from the particle gas.

In practice, $\tilde{g}(e)$ can be calculated in three steps.
First, we solve the HF equations to determine the
self-consistent energies $e^{\rm HF}_i$. In the next step,
we calculate the
positive-energy gas spectrum $e^{\rm HF,free}_i$
{\em at the self-consistent minimum}. In particular, we
take the Coulomb force from the self-consistent calculation.
Finally,
we compute $\tilde{g}_0(e)$ and $\tilde{g}_{\rm free}(e)$ 
using the same folding function.
The quality of approximation (\ref{gapp}) was
tested in Ref.~\cite{[Kru98]} where it was demonstrated that, 
when increasing the number of basis states,
the resulting single-particle level density
quickly converges to the exact result.

\section{Self-consistent Models}\label{models}
\subsection{Skyrme-Hartree-Fock Model}\label{HF}

\noindent
In the SHF method, nucleons are described as nonrelativistic
particles moving independently in a common self-consistent field.
Our implementation of the HF model is based on the standard
ansatz \cite{[Que78]}.
The total binding energy of a nucleus is obtained self-consistently
from the  energy functional:
\begin{eqnarray}
   {\cal E}  =  {\cal E}_{\rm kin}
             &+&{\cal E}_{Sk}
               +{\cal E}_{Sk,ls}\nonumber \\
             &+&{\cal E}_{C}
                +{\cal E}_{\rm pair}
               -{\cal E}_{\text{CM}}, \label{eq:Etot}
\end{eqnarray}
where
$ {\cal E}_{\rm kin}$ is the kinetic energy functional,
$   {\cal E}_{Sk} $ is the Skyrme functional,
$ {\cal E}_{Sk,ls}$ is the spin-orbit functional,
$ {\cal E}_C$ is the Coulomb energy (including the exchange term),
${\cal E}_{\rm pair}$ is the pairing energy, and
${\cal E}_{\text{CM}}$ is the center-of-mass correction.

Since there are more than  80 different Skyrme parameterizations
on the market, the question arises, which forces
should actually be  used when making predictions and comparing with the data?
Here,  we have chosen a small subset of Skyrme forces
which perform well  for  the basic ground-state properties 
(masses, radii, surface thicknesses) and have
sufficiently different properties which allows one to explore the
possible variations among parameterizations.  This subset
contains: SkM$^*$ \cite{[Bar82]}, SkT6
\cite{[Ton84]}, Z$_\sigma$  \cite{[Fri86]},
SkP \cite{[Dob84]}, SLy4
\cite{[Cha95a]}, and SkI1, SkI3, and SkI4 from
Ref.~\cite{[Rei95]}. We have also added the force 
SkO from a recent exploration \cite{[Rei99a]}.
Most of these interactions have been used for the 
investigation of the ground-state properties of SHE before 
\cite{[Cwi99],[Cwi96],[Bur98],[Ben99a],[Rut97]}.
All the selected forces  perform  well concerning 
the total energy and
radii.  They all have comparable incompressibity
$K$=210-250\,MeV and comparable surface energy
which  results from a careful fit to ground-state properties
\cite{[Rei99a]}. Variations occur for properties which are not
fixed precisely by ground-state characteristics. The effective
nucleon mass is 1 for SkT6 and SkP, 0.9 for SkO,
around 0.8 for SkM$^*$ and Z$_\sigma$, and even lower,
around 0.65, for SLy4, SkI1, SkI3, and SkI4.
Isovector properties  also exhibit large variations.
For SkI3 and SkI4, the spin-orbit functional is
given in the extended form of \cite{[Rei95]} which allows a
separate adjustment of isoscalar and isovector spin-orbit
force. The standard Skyrme forces use the particular
combination of isoscalar and isovector terms
which were motivated by the derivation
from a two-body zero-range spin-orbit interaction \cite{[Vau72]}.
(For a detailed discussion of the spin-orbit interaction in SHF
we refer the reader to Refs.~\cite{[Ben99a],[Rei95],[Sha95],[Cha95],[Ons97]}.)

\subsection{Relativistic Mean-Field Model}\label{RMF}

\noindent
In our implementation of the RMF model, nucleons are described as
independent Dirac particles moving in local isoscalar-scalar, 
isoscalar-vector, and isovector-vector mean fields usually 
associated with $\sigma$, $\omega$, and $\rho$ mesons,
respectively \cite{[Rei89a]}. 
These couple to the corresponding local densities of the nucleons
which are bilinear covariants of the Dirac spinors similar to 
the single-particle density of Eq.~(\ref{rhosp}).

The RMF is usually formulated in terms of a covariant Lagrangian;
see, e.g., Ref.~\cite{[Rei89a]}. For our purpose we prefer a formulation 
in terms of an energy functional that is obtained by eliminating 
the mesonic degrees of freedom in the Lagrangian. For a detailed 
discussion of the RMF as an energy density functional theory,  see 
Refs.~\cite{[Spe92a],[Schm95a],[Schm95b],[Schm97a]}. The energy functional 
of the nucleus 
\begin{eqnarray}
\label{eq:RMF:Efunc}
{\cal E}_{\rm RMF}
  =     {\cal E}_{\rm kin} 
& + &   {\cal E}_{\sigma}
      + {\cal E}_{\omega}
      + {\cal E}_{\rho}
      \nonumber \\
& + &   {\cal E}_{\rm C}
      + {\cal E}_{\rm pair}
      - {\cal E}_{\rm CM}
\end{eqnarray}
is composed of the kinetic energy of the nucleons 
${\cal E}_{\rm kin}$, the interaction energies of the
$\sigma$, $\omega$, and $\rho$ fields, and the Coulomb energy 
of the protons ${\cal E}_{\rm C}$. All these are bilinear in the 
nucleonic densities as in the case of non-relativistic models
[cf. Eq.~(\ref{O})]. Pairing correlations are treated 
in the BCS approach employing the same non-relativistic pairing
energy functional ${\cal E}_{\rm pair}$ that is used in the SHF
model. The center-of-mass correction ${\cal E}_{\rm CM}$ is
also calculated in a non-relativistic approximation; see
\cite{[Ben99c]} for a detailed discussion.
The single-particle energies $e_i$ needed to calculate the
shell correction are the eigenvalues of the one-body Hamiltonian 
of the nucleons which is obtained by variation of the energy 
functional (\ref{eq:RMF:Efunc}).

In the context of our study, it is important to note that
the spin-orbit interaction emerges naturally in the RMF
from the interplay of scalar and vector fields \cite{[Rei89a]}. 
Without any free parameters fitted to single-particle data,
the RMF gives a rather good description of spin-orbit splittings 
throughout the chart of nuclei \cite{[Ben99a]}.

As in  SHF, there exist many RMF
parameterizations  which differ in details. 
For the purpose of  the present
study, we choose 
the most successful (or most commonly used)  ones: 
NL1 \cite{[Rei86]}, NL-Z \cite{[Ruf88a]}, NL-Z2 \cite{[Ben99a]}, 
NL-SH \cite{[Sha93]}, NL3 \cite{[Lal97]}, and TM1 \cite{[Sug94a]}. 
All of them
have been used for  investigations of SHE
\cite{[Ben99a],[Rut97],[Lal96b],[Ben99d]}.

The parameterization NL1 is a fit of the RMF along the strategy of 
Ref.~\cite{[Fri86]} used also for the Skyrme interaction Z$_\sigma$. 
The NL-Z parametrization
is a refit of NL1 where the correction for spurious 
center-of-mass motion is calculated from the actual many-body
wave function, while
NL-Z2 is a recent variant of NL-Z with an improved isospin dependence.
The force NL3 stems from a fit including exotic 
nuclei, neutron radii, and information on giant resonances.
The  NL-SH parametrization
was fitted with a bias toward isotopic trends and it
 also uses information on neutron radii. The force TM1 was optimized 
in the same way as NL-SH except for introducing an
additional quartic self-interaction of the isoscalar-vector field
to avoid instabilities of the standard model which occur for 
small nuclei. 
For SHE, the results obtained with NL-Z are not 
distinguishable from results obtained with the parameterization 
PL-40 that is contained in exactly the same manner as NL-Z but uses
a stabilized non-linearity of the scalar-isoscalar field \cite{[Rei88b]}.
(PL-40 was employed in some recent investigations of the
properties of superheavy nuclei \cite{[Bur98],[Rut97],[Ben98a]}.)

All the above  parameterizations provide a good description of 
binding energies, charge radii, and surface thicknesses 
of stable spherical nuclei with the same overall quality as 
the SHF model. The nuclear matter properties of the RMF forces, 
however, show some systematic differences as compared to Skyrme forces. 
All RMF forces have comparable small effective
masses around $m^*/m \approx 0.6$. (Note that the effective mass in 
RMF depends on momentum; hence  the effective mass 
at the Fermi energy is approximately $10\%$ larger.) Compared with
the SHF model, the absolute value of the energy per nucleon is 
systematically larger, with values around $-16.3$ MeV, 
while the saturation density is always 
slightly smaller with typical values around 0.15 nucleons/fm$^{3}$.
The compressibility of the RMF forces ranges from low
values around 170 MeV for NL-Z  to 
$K$=355 MeV for NL-SH, which is rather high.
There are also differences in isovector properties; the symmetry
energy coefficient of all RMF forces is systematically larger than for
SHF interactions, with values between 36.1 MeV for NL-SH and
43.5 MeV for NL1 (see discussion below).

\subsection{Details of Calculations}

\noindent
In order to probe the single-particle shell structure of SHE, 
SHF and RMF calculations were carried out under the assumption of spherical
geometry. By doing so we intentionally disregard deformation effects which
make it difficult to compare different models and parametrizations.
For the same reason, pairing correlations were practically neglected.
(In order to obtain self-consistent spherical 
solutions for open-shell nuclei, small constant pairing gaps,
$\Delta$$<$100\,keV were assumed; the corresponding pairing energies
are negligible. This procedure is approximately equivalent to 
the filling approximation.)
%
%
\begin{figure*}[t!]
\begin{center}
\leavevmode
\epsfxsize=16cm
\epsfbox{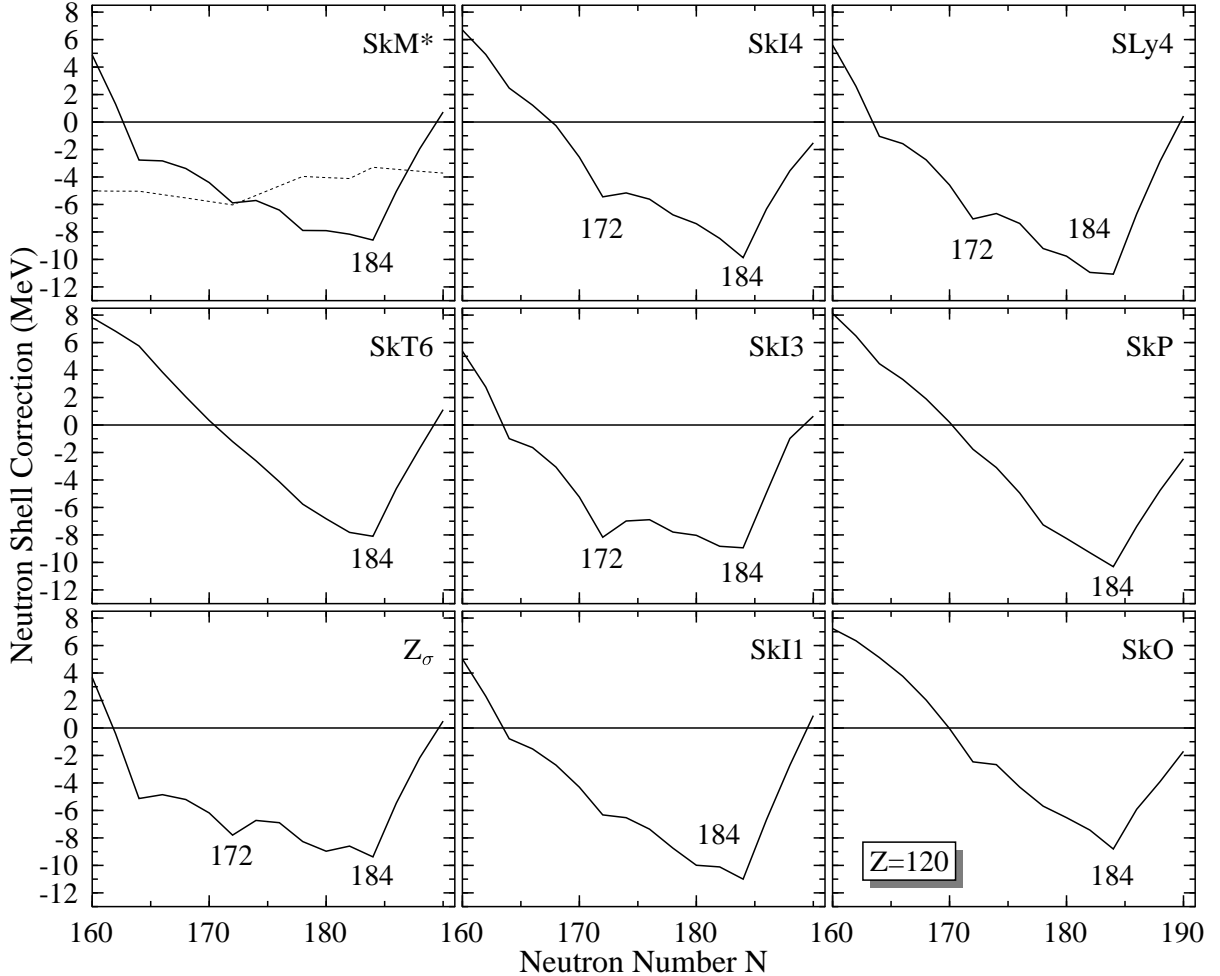}
\end{center}
\caption{Spherical neutron shell corrections for the $Z$=120 isotopes
calculated in nine Skyrme-Hartree-Fock models. The dotted line for 
SkM$^*$ shows the proton shell correction for comparison.
In all nine cases, the minimum of the shell correction is predicted at
$N$=184.
}
\label{shchf1}
\end{figure*}
%
%

The SHF calculations were carried using the coordinate-space
Hartree-Fock code of Ref.~\cite{[Rei91]}. 
The HF equations were solved by the discretization method.
To obtain a proper description of quasi-bound states,
it was necessary to take a very large  box and a very dense mesh.
The actual box size was chosen to be
21\,fm  and the mesh spacing was 0.3\,fm. 
With this choice, the low-lying positive-energy proton states obtained
in SHF perfectly reproduce proton  resonances obtained by solving the
Schr\"odinger equation for the HF potential with purely outgoing boundary
conditions.
 %
%
\begin{figure*}[t!]
\begin{center}
\leavevmode
\epsfxsize=16cm
\epsfbox{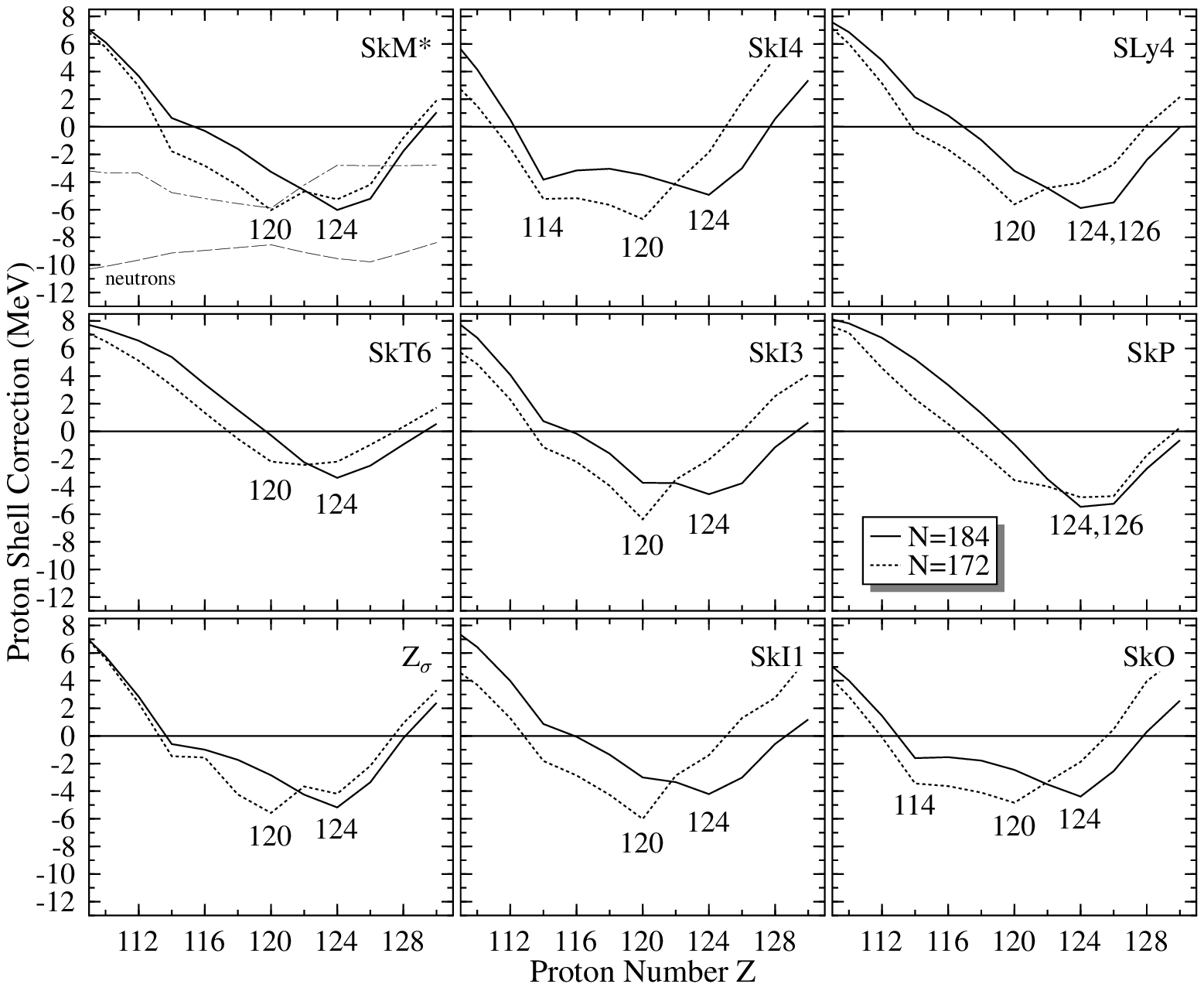}
\end{center}
\caption{Spherical proton shell corrections for
the chains of $N$=184 isotones (solid lines) and
$N$=172 isotones (dotted lines) calculated in nine 
Skyrme-Hartree-Fock models. The dashed (dash-dotted) 
line for SkM$*$  shows the neutron shell correction
for $N$=184 ($N$=172). For $N$=184, the minimum of the shell 
correction is predicted at $Z$=124-126 for all 
parametrizations.
}
\label{shchf2}
\end{figure*}
%
%

The Strutinsky procedure contains two free parameters, the smoothing 
parameter $\gamma$ and the order of the curvature correction $p$.
In calculating the Strutinsky smooth energy, instead of the 
traditional plateau  condition we applied 
the generalized plateau  condition described in Ref. \cite{[Ver98]}.
The optimal values of $\gamma$ (in units of oscillator
frequency $\hbar\omega_0$=41/$A^{1/3}$) calculated for several nuclei
turned out to be close to $\gamma_p$=1.54 and
$\gamma_n$=1.66 for protons and neutrons, respectively;
these values, together with $p$=10,
were adopted in our calculations of shell corrections in SHF.

In the RMF approach, the shell correction can be extracted
from the single-particle spectrum like in SHF. To demonstrate it,
one proceeds along the steps discussed in  Sec.~\ref{SET}. 
The total RMF energy (\ref{eq:RMF:Efunc}) can be decomposed 
into a smooth part and a correction
that fluctuates according to the actual level density.
Since the RMF energy functional is bilinear in the densities,
the extracted shell correction should be accurate up to 
order $O(\delta \rho^2)$.

The RMF calculations were carried out using the coordinate-space
code of Ref.~\cite{[Rut99]}. As in the SHF case, the box size 
was chosen to be 21\,fm with a mesh spacing of 0.3\,fm.

As already mentioned, all successful RMF parameterizations 
give a rather  small effective mass. This leads to a small 
level density around the Fermi surface which in turn requires 
a very large smoothing range $\gamma$ when calculating the smoothed 
level density $\tilde{g}$.
The values for $\gamma$ are strongly correlated with the order 
of the curvature correction polynomial $p$ \cite{[Ver98]}; the 
value $p$=10 chosen here is large enough to provide 
in nearly all cases a sufficiently smooth $\tilde{g}$, but also small 
enough that we can restrict the model space to levels up to 60 MeV, 
which is much larger than the space used in usual RMF calculations.
We have adjusted the smoothing range $\gamma$ to the actual level
density of a large number of nuclei to fulfill a generalized plateau 
condition along the strategy of \cite{[Ver98]}. 
This leads always to values around $\gamma_p$=2.0 for protons and 
$\gamma_n$=2.2 for neutrons. All results presented in this 
paper are calculated with $p$=10 and $\gamma$ fixed at 
these values.

\section{Results}\label{results}
\subsection{Spherical Shell Corrections in Superheavy Nuclei}\label{shcor}

%
%
\begin{figure*}[t!]
\begin{center}
\leavevmode
\epsfxsize=16cm
\epsfbox{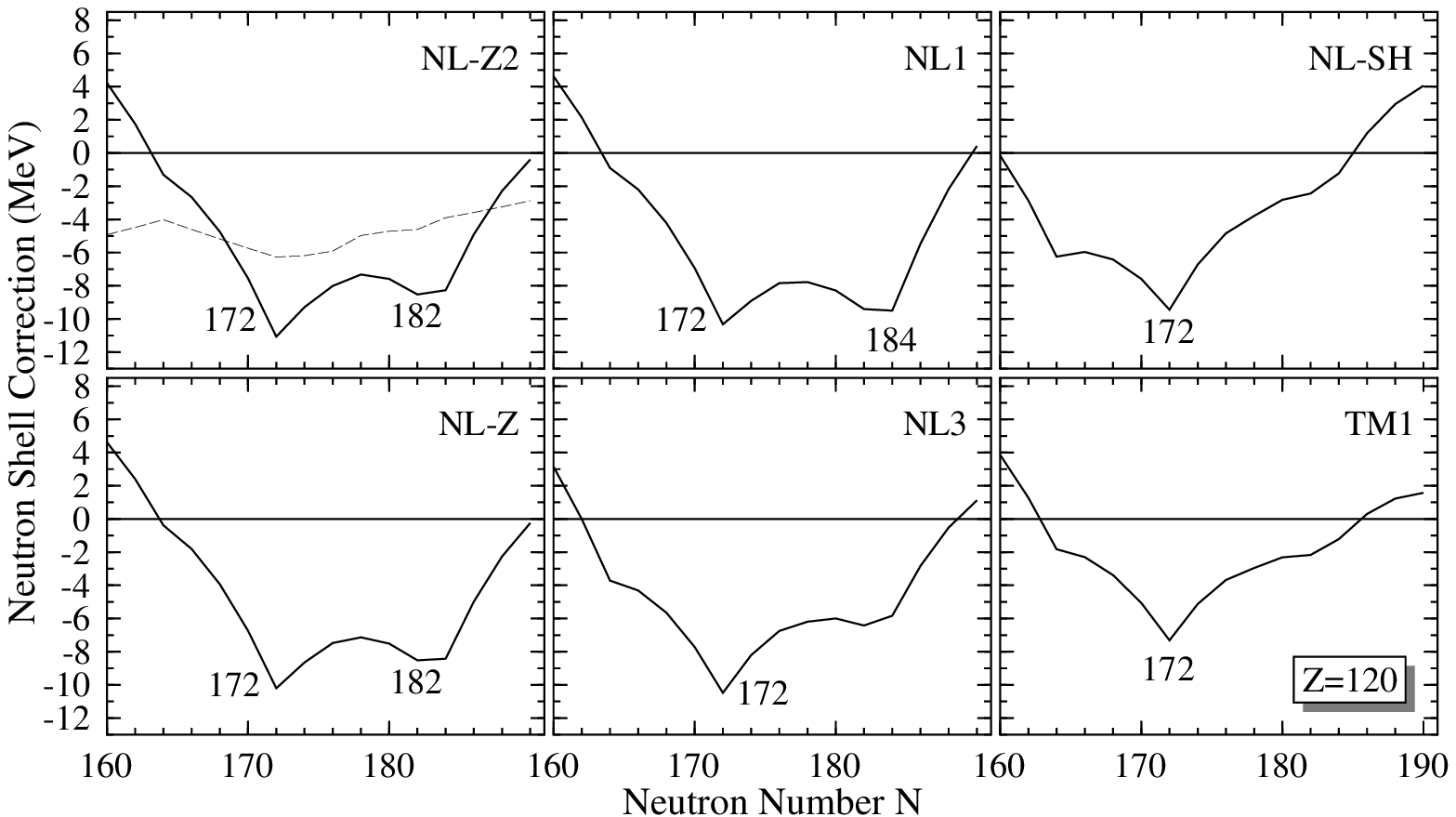}
\end{center}
\caption{Spherical neutron shell corrections for the $Z$=120 isotopes
calculated in six relativistic mean-field models.
In all six cases, the minimum of the shell correction is predicted at
$N$=172. For NL-Z2 the proton shell correction is given 
for comparison by the dashed line.
}
\label{shcrmf1}
\end{figure*}
%
%

\noindent
According to the SHF calculations of Ref.~\cite{[Cwi96]}, the
spherical  magic neutron
number in the SHE region is $N$=184; all the $N$=184 isotones have been 
predicted to have spherical shapes. The magicity of $N$=184
in SHF  is confirmed in this study.
Figure~\ref{shchf1} displays neutron shell
correction calculated in several SHF models
as a function of $N$ for $Z$=120. The absolute
minimum of shell energy always appears
at $N$=184. The $N$=172 shell effect is also seen, but
it exhibits a strong force-dependence  (it is particularly
pronounced for Z$_\sigma$, SkI3, SkI4 and SLy4).

As already mentioned, the neutron levels have the same ordering
for nearly all forces; all differences seen in the shell corrections
are therefore caused by slight changes in the relative
distances of the single-particle levels between the models.
Forces with large effective masses  like SkO,
SkP, and SkT6, give a comparatively large level density
which washes out the shell effects below $N$=184. Forces with
small effective masses (i.e.,  smaller level density)
are much more likely to show significant shell effects at 
lower neutron numbers around $N$=172.

At fixed $Z$, the proton shell correction changes 
rather gradually
as a function of neutron number; this is illustrated in
Fig.~\ref{shchf1} for the Skyrme force SkM$^*$. (Most of the
Skyrme forces give a similar result.) Note that the proton
shell corrections are generally smaller than those 
for the neutrons.  At a second glance, however, one sees that the 
slow variations of the proton shell correction with neutron number 
are correlated with neutron shell closures. 
For instance, the $Z$=120 shell 
correction  is largest 
at  neutron numbers around $N$=172 and it becomes
reduced when approaching  $N$=184. This is caused
by the self-consistent rearrangement of single-particle levels
according to the actual density distribution in the nucleus
and cannot appear in macroscopic-microscopic models with 
assumed average  potentials (see Refs.~\cite{[Ben99a],[Rut97]} 
for more discussion related to this point).

Proton shell corrections for the  $N$=184 and $N$=172 
isotones, obtained in the SHF model, are displayed in Fig.~\ref{shchf2} 
as a function of $Z$. For SkM$^*$,  neutron
shell corrections are also shown
for  the $N$=172 and $N$=184 isotones. 
The shift of the magic proton number with neutron number when going from
$N$=172 to $N$=184 is clearly visible. For $N$=172 most of the Skyrme 
forces (exceptions are SkT6 and SkP) agree on a magic $Z=120$, while for
$N=184$ the shell correction shows a minimum at $Z$=124--126 
in all cases. (Actually, in most cases, shell-corrections 
slightly favor $Z$=124 over $Z$=126; this is related
to the  gradual increase of single-particle energies of 3$p_{3/2}$
and 3$p_{1/2}$ orbitals above $Z$=120.)

Proton shell corrections and the
$N$=172 neutron shell corrections 
are systematically smaller than 
those for neutrons at $N$=184. This partly explains
why
spherical ground states of SHE  are so well
correlated with the magic
neutron number $N$=184, 
 see, e.g., \cite{[Cwi96],[Lal96],[Bur98]}. Note that 
for the majority of Skyrme forces
the $N$=172 isotones are predicted to be deformed.

Skyrme forces with non-standard isospin dependence of the 
spin-orbit interaction are the only ones that give additional 
(but not very pronounced) shell closures. In the 
SkI4 model, there appears a secondary minimum at $Z$=114 for $N$=184, 
while SkI3 is the only Skyrme force which points at $Z$=120 
also for $N$=184. A non-standard spin-orbit interaction, however, 
does not neccesarily lead to shell closures other than $Z$=124-126
for $N$=184.
For SkO, which has   a spin-orbit force 
that is similar to SkI4,  the $Z$=114 shell
is only hinted.
It is to be noted that for several interactions
such as Z$_\sigma$, SkI$x$, and SkO, shell correction changes rather slowly
between $Z$=114 and $Z$=126. This indicates that none of the proton
shell gaps  in this region can be considered as truly ``magic".
(The weak $Z$-dependence of proton shell correction above $Z$=114
was pointed out in the early Ref.~\cite{[Bol71]}.) 
%
%
\begin{figure*}[t!]
\begin{center}
\leavevmode
\epsfxsize=16cm
\epsfbox{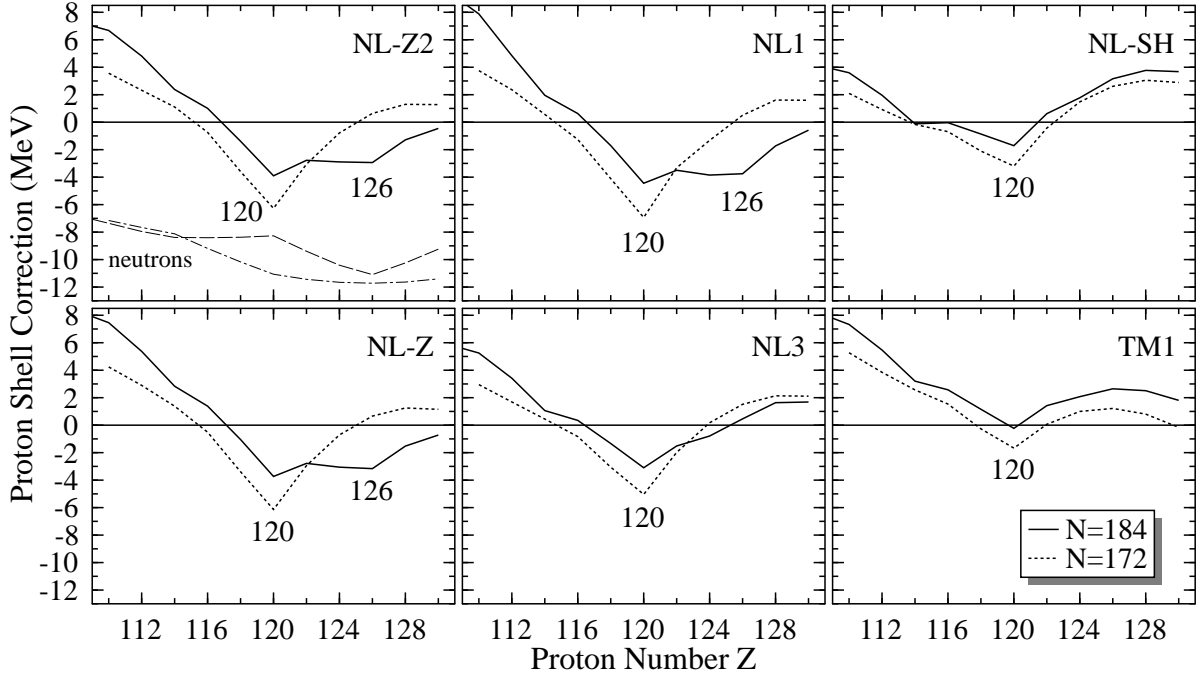}
\end{center}
\caption{Spherical proton shell corrections for
the  $N$=184 (solid line) and $N$=172 (dotted line) isotones
calculated in six relativistic mean-field models.
For all parametrizations, the minimum of the proton shell 
correction is predicted at $Z$=120. For NL-Z2 the neutron
shell correction for the $N$=184 (dashed line) and $N$=172 
(dash-dotted line) isotones are given for comparison.
}
\label{shcrmf2}
\end{figure*}
%
%

The RMF results presented in Figs.~\ref{shcrmf1} and \ref{shcrmf2}
show a pattern that is internally consistent but different from that
of SHF. The minimum of neutron shell correction is systematically
predicted at $N$=172. Except for NL-SH and TM1, the shell effect
at $N$=182-184 is also clearly seen. Note that the $N$=184 gap in the 
single-particle spectrum
is in all cases larger than the one at $N$=182 (see Fig.~\ref{spe2}).
The gaps are separated by a single $4s_{1/2}$ level which  contributes
very weakly to the shell energy. To illustrate
 the variation
of proton shell effects along the $Z$=120 chain,  proton
shell corrections in NL-Z2 are also displayed
in Fig.~\ref{shcrmf1}. Their pattern is very similar to that obtained
in SHF models.

Looking at the proton shell corrections along the chain of $N$=184
isotones, see Fig.~\ref{shcrmf2}, the strongest shell effect 
is now obtained for $Z$=120. When comparing the results for
the $N$=184 and $N$=172 chains, it can be seen again that
the proton shell correction at $Z$=120 is strongly correlated 
with neutron number $N$=172. However,   unlike in  SHF, 
the $Z$=120 shell does not vanish completely for $N$=184.
Proton shell corrections obtained with
NL1, NL-Z, and NL-Z2 at $N$=184 vary rather slowly between $Z$=120 and
$Z$=126, and this resembles the patterm obtained in SHF.
Again, as in the case of Skyrme forces, proton shell corrections
in  RMF are smaller 
than those for the neutrons (cf.  NL-Z2 calculations in
Fig.~\ref{shcrmf2}). 
The increase in the proton shell correction at very
large values of $Z$ for TM1 is related to the spherical $Z$=132 
shell predicted by this interaction \cite{[Rut97]}.

Shell closures can also be analysed in terms of the 
two-neutron and two-proton shell gaps
\begin{eqnarray}\label{deltas}
\delta_{2n} & = & E(N+2,Z)-2E(N,Z)+E(N-2,Z), \nonumber\\
\delta_{2p} & = & E(N,Z+2)-2E(N,Z)+E(N,Z-2) 
\end{eqnarray}
discussed in Refs.~\cite{[Rut97],[Ben99d]}. 
The pattern of shell corrections calculated in SHF
and RMF nicely follows the behavior of neutron and proton
shell gaps found there. In particular, the strong
correlation between shell effects at $Z$=120 and $N$=172 
in RMF is seen in both representations. While shell gaps
are related (but not equivalent) to the gaps in the 
single-particle spectrum, the shell correction gives 
also a measure of the stabilizing effect of a shell closure
on the nuclear binding energy.

\subsection{Macroscopic Energies}

\begin{figure*}[t!]
\begin{center}
\leavevmode
\epsfxsize=14cm
\epsfbox{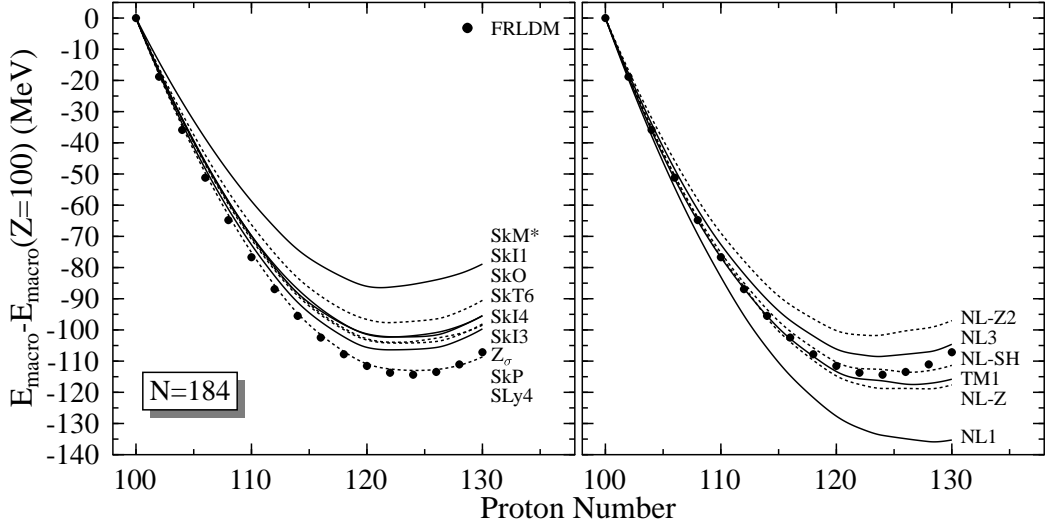}
\end{center}
\caption{Macroscopic energy $\tilde{E}^{\rm HF}$
(\protect\ref{Emacro}) extracted from the calculated Hartree-Fock
energies of the $N$=184 isotones. For comparison, the
phenomenological 
macroscopic energy of the Yukawa-plus-exponential mass formula (FRLDM)
of Ref.~\protect\cite{[Mol88]} with parameters of
Ref.~\protect\cite{[Cwi94a]} is also displayed. 
To illustrate $Z$-dependence, all
energies were normalized to the FRLDM value at $Z$=100.
}
\label{macrohf}
\end{figure*}

\noindent
By subtracting the shell correction from the calculated
binding energy, one obtains a rough estimate for
the associated macroscopic energy $E_{\rm macro}$
(\ref{Emacro}). The macroscopic part of the SHF and RMF energies 
for the $N$=184 isotones as a function of $Z$ 
is displayed in Fig.~\ref{macrohf}. 
The macroscopic energy of the Yukawa-plus-exponential mass formula
of the Finite-Range Liquid Drop Model  (FRLDM)
of Ref.~\protect\cite{[Mol88]}, with parameters of
Ref.~\protect\cite{[Cwi94a]}, is also shown for comparison. 
To illustrate $Z$-dependence, all
energies were normalized to the value at $Z$=100.
In general, the behavior of $E_{\rm macro}$ is similar in
all cases. In particular, the macroscopic proton drip 
line is consistently predicted to be at $Z$$\approx$120-124. 
It is interesting  to note that the only Skyrme force 
which agrees with FRLDM is SLy4; other forces deviate 
from it significantly. 
The RMF forces give qualitatively the same results; there
are several forces (NL-Z, TM1, and NL-SH) which give 
values of $E_{\rm macro}$ close to the FRLDM.
%
%
%
%

\begin{figure*}[t!]
\begin{center}
\leavevmode
\epsfxsize=14cm
\epsfbox{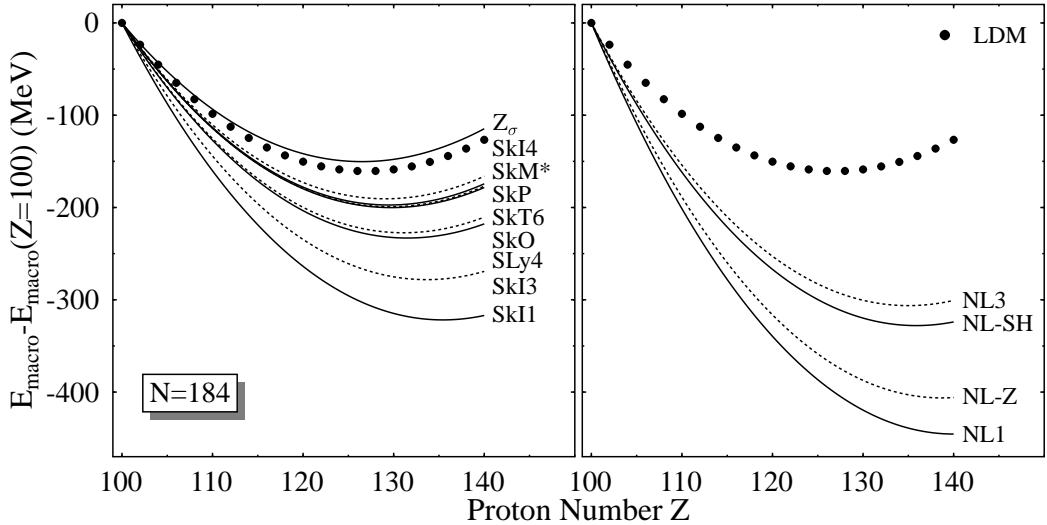}
\end{center}
\caption{Macroscopic energy $E_{\rm macro,LDM}$ (\protect\ref{LDM})
for the $N$=184 isotones as a function of $Z$ calculated
for several Skyrme forces (lines) and the standard 
liquid drop model (dots). To illustrate the $Z$ dependence, all
energies were normalized to zero at $Z$=100. The
bulk parameters of Skyrme forces
are given in Table~\protect\ref{table1}. The Coulomb-energy
constant was assumed to be $a_{\rm Coul}$=0.717 MeV
\protect\cite{[Mye69]} in all cases.
}
\label{ldp}
\end{figure*}

In an attempt to understand
the pattern shown in Fig.~\ref{macrohf}, we employed the simple
liquid drop model  expression
\begin{eqnarray}\label{LDM}
E_{\rm macro,LDM} = a_{\rm vol}  A & + & a_{\rm surf} A^{2/3}
\\ \nonumber & + & 
a_{\rm sym} \frac{(N-Z)^2}{A}
   +  a_{\rm Coul} \frac{Z^2}{A^{1/3}}.
\end{eqnarray}
The parameters $a_i$ of Skyrme and RMF forces were calculated 
in the limit of symmetric nuclear matter; 
they are given in Table~\ref{table1},
together with the values for  the standard liquid drop
model (LDM) of Ref.~\cite{[Mye69]}. [Note that these 
values slightly change when including higher-order
terms in the LDM expansion (\ref{LDM}).]
Figure~\ref{ldp} shows the macroscopic energy
(\ref{LDM}) as a function of $Z$ for the $N$=184 isotones.
The huge differences between results for various Skyrme
and RMF parametrizations can be traced back to their different 
symmetry-energy coefficients. Indeed, for most of the forces 
discussed, $a_{\rm sym}$ is significantly greater than that of LDM,
and this results in an increased slope of $E_{\rm macro, LDM}$.
For the RMF forces the significantly larger $a_{\rm vol}$
even further increases the difference with respect to the LDM. 
Unfortunately, there is very little similarity between
the results of microscopic calculations of Fig.~\ref{macrohf}
and the results of expansion (\ref{LDM}). When comparing
the energy scales of Figs.~\ref{macrohf} and~\ref{ldp}, 
one finds huge differences, of the order of 100 MeV,  between
$E_{\rm macro}$ and $E_{\rm macro, LDM}$. While for
RMF   the energy ordering remains the same
in both cases, this feature does not hold for SHF.
Only when looking at $E_{\rm macro, LDM}$, the results
are ordered according to the corresponding
 values of $a_{\rm sym}$, 
as expected. All this indicates
that even for very heavy nuclei with $A$$\sim$300, the simple
leptodermous expansion with parameters taken from
nuclear matter calculations is not going to work 
\cite{[Bra85],[Bra97]}; the finite-size effects are still
very important for SHE.
%
%
\begin{table}[b!]
\caption{Key properties of symmetric nuclear matter for the
Skyrme and RMF forces used in this paper:
binding energy per nucleon, surface energy, and
symmetry energy, all in MeV. The RMF values for 
$a_{\rm surf}$ are taken from Ref.~\protect\cite{[Sto98]}.
The standard liquid drop
model (LDM) values \protect\cite{[Mye69]} are also shown.
}
\begin{tabular}{lccc}
Force & $a_{\rm vol}$ & $a_{\rm surf}$  & $a_{\rm sym}$  \\ 
\hline
  SkM$^*$    &  $-15.9$ &  17.59 & 30.0 \\
  Z$_\sigma$ &  $-15.9$ &  16.94 & 26.7 \\
  SkT6       &  $-16.1$ &  18.12 & 29.9 \\
  SLy4       &  $-16.1$ &  18.18 & 32.0 \\
  SkI1       &  $-15.9$ &  17.31 & 37.5 \\
  SkI3       &  $-16.0$ &  17.52 & 34.8 \\
  SkI4       &  $-15.9$ &  17.28 & 29.5 \\
  SkP        &  $-16.0$ &  17.95 & 30.0 \\
  SkO        &  $-15.8$ &  17.00 & 32.0 \\[1mm]
  NL1        &  $-16.4$ &  18.66 & 43.5 \\
  NL-Z       &  $-16.2$ &  17.72 & 41.7 \\
  NL-Z2      &  $-16.1$ &        & 39.0 \\
  NL3        &  $-16.2$ &  18.46 & 37.4 \\
  NL-SH      &  $-16.3$ &  19.05 & 36.1 \\
  TM1        &  $-16.3$ &        & 36.9 \\[1mm]  
  LDM        &  $-15.7$ &  18.56 & 28.1  
\end{tabular}
\label{table1}
\end{table}
%
%

In spite of the fact that macroscopic energies extracted from
different self-consistent models  systematically differ,
the corresponding shell corrections are similar. For instance, 
the general pattern and magnitude of shell energies displayed in 
Figs.~\ref{shchf1} and \ref{shchf2}
do not depend very much on the Skyrme interaction used, and the same is
true for the RMF results shown in Figs.~\ref{shcrmf1} and \ref{shcrmf2}.
This means that although the global properties of effective interactions
employed in this work differ, their single-particle spectra are fairly
similar. Hence, shell corrections extracted
from self-consistent single-particle spectra are very useful measures of
spectral properties of effective forces.
Figure~\ref{macrohf} also illustrates 
how dangerous it is to extrapolate self-consistent results
in the region of SHE. The trends of relative 
 binding energies
(e.g., $Q_\alpha$ values) are expected
to smoothly deviate from force to force. The nice agreement
with experimental data for the heaviest elements
obtained in the SHF calculations with
SLy4 \cite{[Cwi99]} and  in the macroscopic-microscopic
calculations with the FRLDM \cite{[Cwi94a]} 
indicates that the macroscopic energies of 
forces which are too far off the FRLDM values, i.e.\
SkM$^*$, SkI1, and NL1, are probably not reliable in this region.
%
%
%
%

Figure~\ref{macrohf} shows  that the  power
of a force for predicting  total binding energies 
is fairly independent of its predictive power for shell 
effects. Forces with a similar (good) description of the
smooth trends of binding energies can yield  rather different
magic numbers; compare, e.g., SLy4 and NL3.

\section{Conclusions}\label{conclusions}

\noindent
The recent experimental progress in the 
search for new superheavy
elements opens a new window for systematic explorations of the
limit of nuclear mass and charge. Theoretically, predictions
in the region of SHE are bound to be extrapolations from the
lighter systems. An interesting and novel feature of SHEs is
that the Coulomb interaction can no longer be treated as a small
perturbation atop the nuclear mean field; its feedback on
the nuclear potential is significant.

The main objective of this study was to perform a detailed analysis
of shell effects in SHE. Since many nuclei from this region are
close to the proton drip line, a new method of calculating
shell corrections, based on the Green's function
approach,  had to be developed. This technique was applied to a
family of Skyrme interactions and to several
RMF parametrizations. This tool turned out to be extremely useful
for analyzing the spectral properties of self-consistent
mean fields.

It has been concluded that both the SHF and RMF calculations are
{\em internally} consistent. That is, {\em all} the Skyrme
models employed in this work predict 
the strongest spherical shell effect at
$N$=184 and $Z$=124,126.
On the other hand, {\em all} the
RMF parametrizations yield the strongest shell
effect at $N$=172 and $Z$=120. It is very likely that
the main factor contributing to this difference is the
spin-orbit interaction, or rather its isospin dependence
\cite{[Ben99a],[Rei95],[Sha95],[Cha95],[Ons97]}. The role of 
the spin-orbit potential in determining the stability of SHE
was posed already in the seventies  \cite{[Mos78],[Tan79]}.
The experimental determination of the centre of
shell stability in the region of SHE will, therefore, be of
extreme importance for pinning down the question of the
spin-orbit force.

Another interesting conclusion of our work is that the
pseudo-spin symmetry seems to be strongly violated in the
RMF calculations for SHE. As a matter of fact,
the $N$=172 and $Z$=120
magic gaps predicted in the relativistic model 
appear as a direct consequence of pseudo-spin breaking.
This is quite surprising in light of several recent
works on the pseudo-spin conservation in
RMF \cite{[Gin98],[Men98]}. 

Finally, from  calculated masses
we extracted  self-consistent macroscopic energies. They
show a significant spread when extrapolating to unknown SHE.
This is expected to give rise to systematic (smooth) deviations
between masses and mass differences obtained in various 
self-consistent models.

\acknowledgments
This research was supported in part by
the U.S. Department of Energy
under Contract Nos.\ DE-FG02-96ER40963 (University of Tennessee),
DE-FG05-87ER40361 (Joint Institute for Heavy Ion Research),
DE-FG02-97ER41019 (University of North Carolina),
DE-AC05-96OR22464 with Lockheed Martin Energy Research Corp.\ (Oak
Ridge National Laboratory), the Polish Committee for
Scientific Research (KBN) under Contract No.~2~P03B~040~14,
NATO grant CRG 970196, and Hungarian OTKA Grant No. T026244.

\end{document}